\documentclass[12pt]{iopart}
\pdfoutput=1
\usepackage{graphicx}
\begin{document}

\title[Region-graph partition function expansion]{Partition
Function Expansion on Region-Graphs and Message-Passing Equations}

\author{Haijun Zhou$^{1}$, Chuang Wang$^{1}$, Jing-Qing Xiao$^{1,2}$,
and Ze-Dong Bi$^{1}$}

\address{$^1$ State Key Laboratory of Theoretical Physics,
	    Institute of Theoretical Physics, Chinese Academy of Sciences,
		Beijing 100190,	China}

\address{$^2$ Institute of Applied Mathematics, Academy of Mathematics and
Systems Science, Chinese Academy of Sciences, Beijing 100190, China}

\ead{zhouhj@itp.ac.cn}

\begin{abstract}
Disordered and frustrated graphical systems are ubiquitous in physics,
biology, and information science. For models on complete graphs or
random graphs, deep understanding has been achieved through the mean-field
replica and cavity methods. But finite-dimensional `real' systems persist to
be very challenging because of the abundance of short loops and strong local
correlations. A statistical mechanics theory is constructed in this paper for
finite-dimensional models based on the mathematical framework of
partition function expansion and the concept of region-graphs.
Rigorous expressions for the free energy and grand free energy are derived.
Message-passing equations on the region-graph, such as belief-propagation and
survey-propagation, are also derived rigorously.
\end{abstract}

\pacs{05.50.+q, 02.50.-r, 75.10.Hk, 89.70.-a}

\submitto{JSTAT}

\maketitle


Graphical models are used to describe systems composed of elements which lack
translational degrees of freedom but have changeable internal states.
Disordered and frustrated graphical models are ubiquitous in physics
(spin glasses), biology (gene regulation and neuron networks), and
information science (error-correcting codes, constraint satisfaction problems).
Revealing and understanding the rich and fascinating properties of these
systems has been a major field of statistical physics for many years
\cite{Mezard-etal-1987,Mezard-Montanari-2009}. On the theoretical side, a deep
understanding of the equilibrium behaviors of models defined on complete
graphs (e.g., the Sherrington-Kirkpatrick model) or  finite-connectivity
random graphs (e.g., the Viana-Bray model) has been achieved by the mean-field
replica and cavity methods
\cite{Mezard-etal-1986,Monasson-1995,Mezard-Parisi-2001,Mezard-etal-2002,Krzakala-etal-PNAS-2007}.

Due to the finite-dimensional nature, real-world graphical model systems (e.g.,
the Edwards-Anderson (EA) model \cite{Edwards-Anderson-1975}) usually contain
extremely many short loops with strong interaction strengths, which cause
strong local correlations within groups of elements. The abundance of short
loops makes theoretical study of finite-dimensional graphical models very
difficult, and so far progress has mainly been achieved through extensive
numerical simulations (see, e.g., \cite{Belletti-Cotallo-Cruz-etal-2008} and
references therein). The high-temperature equilibrium properties of
finite-dimensional models can be studied by the cluster variation method
\cite{Kikuchi-1951,An-1988} and its recent extension, the region-graph
method \cite{Yedidia-Freeman-Weiss-2005}. But for the most interesting
low-temperature regime, a powerful and general theoretical framework is still
lacking. Recently, Rizzo and co-workers \cite{Rizzo-etal-2010} made an initial
step in combining the cluster variation method with the replica method to
study the low-temperature behavior of the EA model, but several technical
difficulties of this approach still need to be remedied (such as the
issue of non-positive-definite message functionals).

In this letter, a statistical mechanics theory for finite-dimensional graphical
models is rigorously constructed  based on the mathematical framework of
partition function expansion \cite{Xiao-Zhou-2011,Chertkov-Chernyak-2006b} and
the region-graph concept of Yedidia and co-workers
\cite{Yedidia-Freeman-Weiss-2005}. In this theoretical framework, the sets of
elements that form local loops are regarded as regions, and the possible strong
local correlations within the regions are explicitly taken into account.
Free energy expressions and region-graph message-passing equations at
different hierarchies of the free energy landscape are derived without using
any approximations. The theoretical framework directly corresponds to a
distributed algorithm for single graphical systems. This work may find broad
applications in finite-dimensional spin-glass systems and in real-world
information systems (such as image processing).

\begin{figure}
\begin{center}
\includegraphics[width=0.25\linewidth, bb= 10 10 520 520]{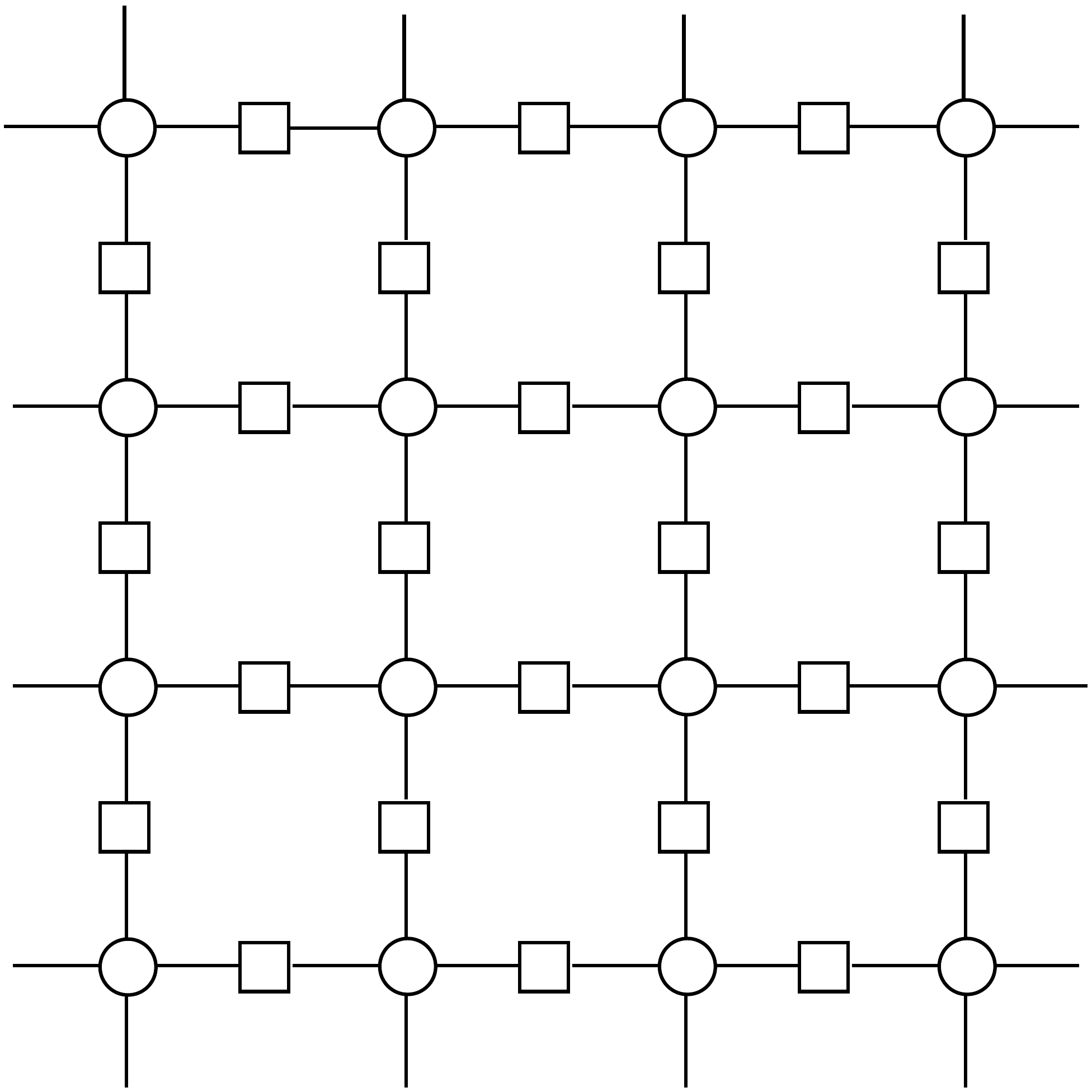}
\end{center}
\caption{                                               \label{fig:factorgraph}
Part of the factor-graph for the two-dimensional Edwards-Anderson model on a
square lattice. Circles represents variable nodes and squares represent
interactions.}
\end{figure}

{\em Region-graph representation}---We consider a general model system defined on
a factor-graph \cite{Kschischang-etal-2001} of $N$ variable nodes
($i=1, 2, \ldots, N$, representing the elements) and $M$ function nodes
($a=1, 2, \ldots, M$,
representing the interactions).
The total-energy function has the following additive form:
\begin{equation}                                         \label{eq:totalenergy}
H(\underline{x}) = \sum\limits_{a=1}^{M} E_a(\underline{x}_{\partial a}) .
\end{equation}
A configuration $\underline{x}$ of the system is defined by the states of all
the $N$ variable nodes,
$\underline{x} \equiv \{x_1, x_2, \ldots, x_N\}$, $x_i$ being the discrete
state of  node $i$. Each function node $a$ is connected to a set $\partial a$
of variable nodes, its energy $E_a$ depends only on the states
$\underline{x}_{\partial a} \equiv \{x_i| i\in \partial a\}$ of the variable
nodes in $\partial a$.
As a simple example, figure~\ref{fig:factorgraph} shows the
factor-graph for the EA model on a two-dimensional (2D) square
lattice, whose energy function is defined as
\begin{equation}
\label{eq:EAmodel}
H(\underline{x}) =
\sum\limits_{\langle i, j\rangle} J_{i j} x_i x_j ,
\end{equation}
where $\langle i, j\rangle$ denotes an edge between two nearest-neighboring
lattice sites $i$ and $j$,  $J_{i j}$ is the coupling constant on this
edge (the state of each variable node is binary in the EA model,
$x_i \in \{-1,+1\}$). The equilibrium partition function of the general system
\eref{eq:totalenergy} is defined as
\begin{equation}                                  \label{eq:partitionfunction}
    Z(\beta) \equiv \sum_{\underline{x}} e^{-\beta H(\underline{x})}
  = \sum\limits_{\underline{x}} \prod\limits_{a=1}^{M}
     \psi_a(\underline{x}_{\partial a}) ,
\end{equation}
where $\beta$ is the inverse temperature, and the nonnegative function
$\psi_a(\underline{x}_{\partial a})
\equiv e^{-\beta E_a(\underline{x}_{\partial a})}$
is the Boltzmann factor of function node $a$.

The factor-graph of model (\ref{eq:totalenergy}) might contain many short
loops, which cause strong local correlations among the variable nodes. To
better describe the correlations and fluctuations caused by short loops in the
system, it is desirable to view the factor-graph as a union of different
regions. Each region contains a set of variable nodes and function nodes. The
variable nodes that appear in the same region are supposed to be strongly
correlated and are treated with special care. More precisely, a region-graph
$R$ is formed by regions $\{\alpha, \gamma, \mu, \ldots\}$ and the edges
$\{(\alpha, \gamma), (\mu,\nu), \ldots\}$ between regions
\cite{Yedidia-Freeman-Weiss-2005}. A region $\alpha$ contains some variable
nodes and function nodes of the factor graph, it satisfies the condition that,
if a function node $a\in \alpha$, then all the variable nodes connected to $a$
are in $\alpha$ (that is, $\partial a \subset \alpha$). The region concept is
an extension of the cluster concept of the cluster variation method
\cite{Kikuchi-1951,An-1988}. The edges in a region-graph are directed. If there
is an edge pointing from a region $\mu$ to another region $\nu$, then $\nu$
must be a subregion of $\mu$ (namely, if variable node $i \in \nu$, then
$i\in \mu$, and  if  function node $a\in \nu$, then $a\in \mu$).
If there is a directed path from region $\alpha$ to region $\gamma$, then we
say that $\gamma < \alpha$ and $\alpha > \gamma$ (notice that if there is no
directed path from $\alpha$ to $\gamma$, such an ordering relationship does
not exist, even if $\gamma \subset \alpha$).

Each region $\alpha$ of the region-graph has a counting number $c_\alpha$
that is recursively determined by
\begin{equation}                                              \label{eq:rgcn}
c_\alpha = 1 - \sum\limits_{\{\gamma | \gamma \in R, \gamma > \alpha\}}
c_\gamma .
\end{equation}
Notice that $\sum_{\gamma \geq \alpha} c_\gamma =1$ for any region
$\alpha \in R$, that is, each region is counted exactly once
(see \cite{Yedidia-Freeman-Weiss-2005,An-1988}).
The region-graph $R$ and its associated counting numbers are required to
satisfy the following region-graph condition \cite{Yedidia-Freeman-Weiss-2005}:
For any variable node $i$, the subgraph of $R$ formed by all the regions
$\alpha$ containing $i$ and all the edges between these regions is connected;
furthermore,
\begin{equation}
\label{eq:cn}
        \sum\limits_{\{\alpha| i\in \alpha\}}  c_\alpha   = 1 , \quad\quad
        \sum\limits_{\{\alpha | a\in \alpha\}} c_\alpha   = 1 .
\end{equation}
\begin{figure}
\begin{center}
\includegraphics[width=0.4\linewidth, bb= 20 20 575 575]{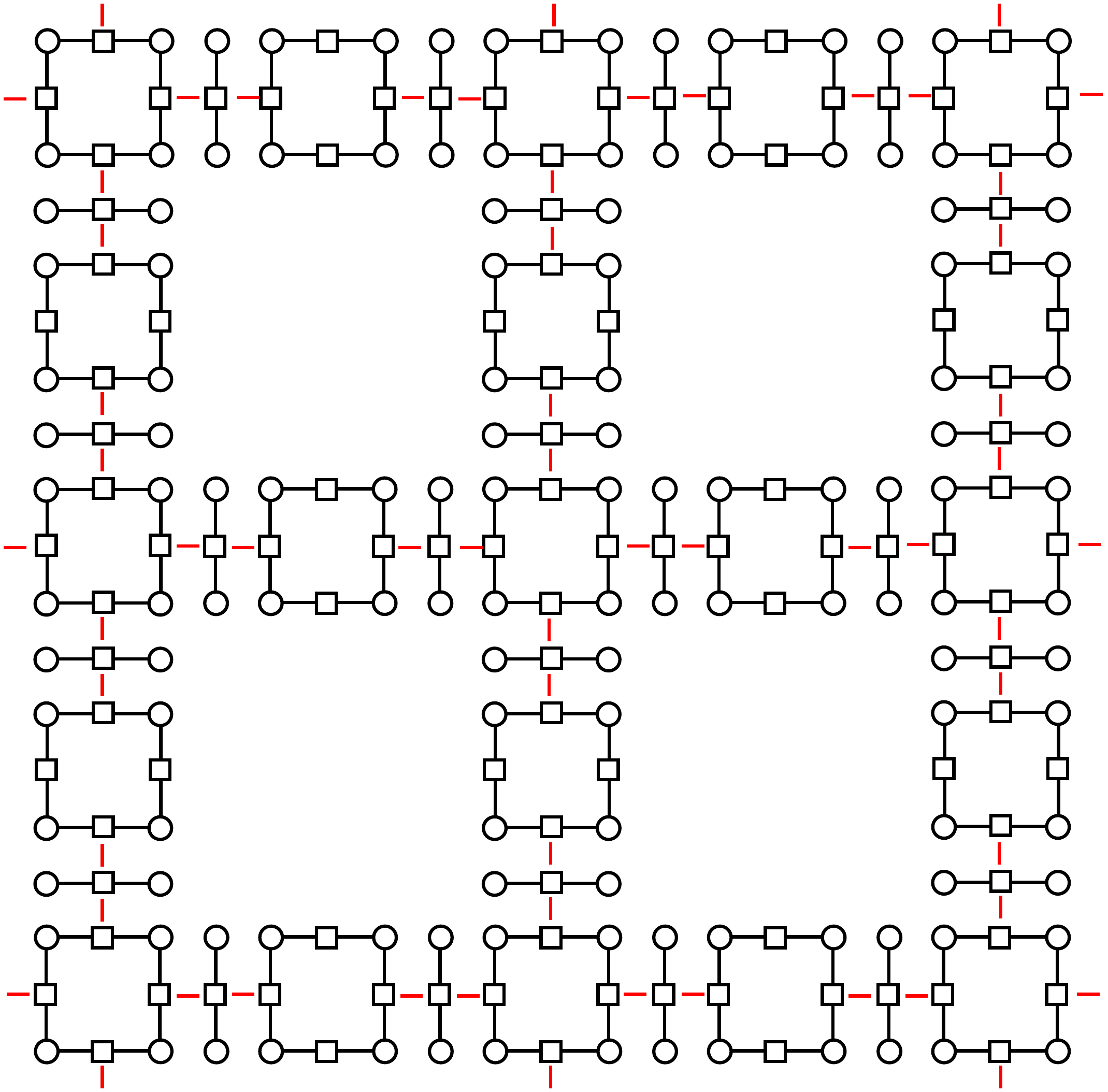}
\end{center}
\caption{                                               \label{fig:regiongraph}
Part of a region-graph for the 2D Edwards-Anderson model, whose factor-graph
is shown in \fref{fig:factorgraph}. There are two types of regions, the square
regions (containing four variable nodes and four function nodes) and the pair
regions (containing two variable nodes and one function node). The red short
lines represent the edges of the region-graph. Each square region has a
counting number $c=1$, and each pair region has a counting number $c=-1$.}
\end{figure}

For the same factor-graph it is usually possible to construct many different
region-graphs which all satisfy the region-graph condition. The factor-graph
itself can be turned into the simplest region-graph with two types of regions
(each small region is formed by a single variable node, and each large region
is formed by a function node and all the connected variable nodes).
Constructing an appropriate region-graph for a graphical model
\eref{eq:totalenergy} is by no means a trivial issue, a lot of physical
insights are needed here. We leave this construction problem as an open issue
for future study, and simply assume a suitable region-graph $R$ for model
(\ref{eq:totalenergy}) is ready for use. To give a simple example, we show in
\fref{fig:regiongraph} part of a region-graph for the EA model
\eref{eq:EAmodel} on a 2D square lattice.

{\em Partition function expansion}---Because of the region-graph condition (\ref{eq:cn}),
we can reformulate the partition function (\ref{eq:partitionfunction}) as
\begin{equation}                                            \label{eq:Z-exp-1}
    Z(\beta) = \sum\limits_{\underline{x}}
    \prod\limits_{\alpha \in R} \biggl[
    \prod_{a\in \alpha} \psi_a(\underline{x}_{\partial a})
    \biggr]^{c_\alpha} .
\end{equation}
For each region $\alpha$ we define its configuration as $\underline{x}_\alpha
\equiv \{x_i^\alpha | i \in \alpha\}$, where $x_i^\alpha$ is the state of
 node $i$ in $\alpha$. For a function node $a$ in region $\alpha$, the
variable states at its neighborhood is then denoted as
$\underline{x}_{\partial a}^\alpha$, with
$\underline{x}_{\partial a}^{\alpha} \equiv
\{ x_i^\alpha | i \in \partial a\}$.
The Boltzmann factor for  region $\alpha$ is denoted as
$\Psi_\alpha(\underline{x}_\alpha)  \equiv \prod_{a\in \alpha}
\psi_a^{c_\alpha}(\underline{x}_{\partial a}^\alpha)$.

A variable node $i$ may belong to two or more regions (say $\alpha, \gamma,
\ldots$). If this is the case, we regard the states
$x_i^\alpha, x_i^\gamma, \ldots$ of node $i$ in the different regions as
different variables. With these auxiliary states
$\{\underline{x}_\alpha | \alpha \in R\}$, we then have
\begin{eqnarray}
    Z(\beta) &=  \sum\limits_{\underline{x}}
    \sum\limits_{\{\underline{x}_\alpha\}}
    \prod\limits_{\alpha \in R} \Bigl[ \Psi_\alpha(\underline{x}_\alpha) \Bigr]
    \prod\limits_{\gamma \in R} \prod_{i \in \gamma} \delta(x_i^\gamma, x_i)
 \label{eq:Z-exp-2} \\
     &=
    \sum\limits_{\{\underline{x}_\alpha\}}
    \prod\limits_{\alpha \in R} \Bigl[ \Psi_\alpha(\underline{x}_\alpha)
   \Bigr] \prod_{(\mu, \nu)\in R}
   \biggl[ \prod_{i\in \mu\cap\nu} \delta(x_i^\mu, x_i^\nu) \biggr] ,
   \label{eq:Z-exp-3}
\end{eqnarray}
where the Kronecker symbol $\delta(x, y)=1$ if $x=y$ and $\delta(x,y)=0$
otherwise. The equivalence of (\ref{eq:Z-exp-3}) and (\ref{eq:Z-exp-2})
is ensured by the region-graph condition that the subgraph induced by any
variable node $i$ is connected. In \eref{eq:Z-exp-3}, the product over all the
edges $(\mu,\nu)$ of the region-graph $R$ makes sure that, the partition
function $Z$ is contributed only by those region configurations for which
each variable node $i$ takes the same state in all the different regions that
contain $i$.  These are the edge constraints on $R$. We notice that, if in $R$
there are two or more directed paths from a region $\alpha$ to another region
$\gamma$, then some of the region-graph edge constraints are redundant in the
sense that they can be removed from the edge product without affecting the
equivalence of \eref{eq:Z-exp-3} and \eref{eq:Z-exp-2}. It is desirable for
a region-graph $R$ to be free of redundant edges. If some edges of $R$ are
redundant, it should not be included in the edge product of \eref{eq:Z-exp-3}.
In what follows, for simplicity, we assume that the region-graph $R$ contains
no redundant edges.

For each edge $(\mu, \nu) \in R$ we introduce two auxiliary probability
distributions $p_{\mu\rightarrow \nu}(\underline{x}_{\nu\cap\mu})$ and
$p_{\nu \rightarrow \mu}(\underline{x}_{\mu\cap\nu})$. They are at the moment
arbitrary functions under the minimal constraints of being nonnegative and
normalized. The state vector $\underline{x}_{\mu\cap\nu}$ is defined as
$\underline{x}_{\mu\cap\nu} \equiv \{x_i^\mu| i\in \mu\cap\nu\}$, similarly
$\underline{x}_{\nu\cap\mu} \equiv \{x_i^\nu| i\in \mu\cap\nu\}$. We then get
the following formula of central importance
\begin{equation}                                             \label{eq:Z-exp-4}
\fl
\quad\quad
    Z(\beta) = \sum\limits_{\{\underline{x}_\alpha\}}
    \prod\limits_{\alpha \in R} \biggl[ \Psi_\alpha(\underline{x}_\alpha)
    \prod_{\gamma \in \partial \alpha}
     p_{\gamma\rightarrow \alpha}(\underline{x}_{\alpha\cap\gamma}) \biggr]
    \prod_{(\mu, \nu)\in R} \biggl[
    \frac{\prod_{i\in \mu\cap\nu} \delta(x_i^\mu, x_i^\nu)}
    {
    p_{\mu\rightarrow \nu}(\underline{x}_{\nu\cap\mu})
    p_{\nu\rightarrow \mu}(\underline{x}_{\mu\cap\nu})} \biggr] ,
\end{equation}
where $\partial \alpha$ denotes the set of regions that are connected with
region $\alpha$ by an edge. A loop series expression for the partition function
(\ref{eq:Z-exp-4}) can be easily written down following \cite{Xiao-Zhou-2011}.
The equilibrium free energy $F(\beta)\equiv -(1/\beta) \ln Z(\beta)$ of the
system is expressed as
\begin{equation} \label{eq:feexpan}
F(\beta) = F_{0} - \frac{1}{\beta} \ln \biggl[1+ \sum\limits_{r_{loop}
\subseteq R} L_{r_{loop}} \biggr] ,
\end{equation}
where $r_{loop}$ denotes any subgraph that is free of dangling edges (each
$\alpha \in r_{loop}$ is connected with at least two other regions of
$r_{loop}$), and $L_{r_{loop}}$ is its correction contribution to the free
energy. To ensure that any subgraph $r$ with at least one dangling edge
has zero correction contribution, the auxiliary probability functions need to
satisfy the following equation
\begin{equation} \label{eq:rgbp}
\fl \quad\quad
    p_{\mu\rightarrow \nu}(\underline{x}_{\nu\cap\mu})
    =  B_{\mu\rightarrow \nu}(\{p_{\gamma\rightarrow \mu}\})
    \equiv
    \frac{
    \sum\limits_{\underline{x}_\mu}
    \prod\limits_{i\in \mu\cap\nu} \delta(x_i^\mu, x_i^\nu)
    \Psi_\mu(\underline{x}_\mu)
    \prod\limits_{\gamma \in \partial \mu\backslash \nu}
    p_{\gamma\rightarrow \mu}(\underline{x}_{\mu\cap\gamma})}
    {
    \sum_{\underline{x}_\mu}
    \Psi_\mu(\underline{x}_\mu)
    \prod_{\gamma \in \partial \mu\backslash \nu}
    p_{\gamma\rightarrow \mu}(\underline{x}_{\mu\cap\gamma})} .
\end{equation}
This self-consistent equation is referred to as the region-graph
belief-propagation (rgBP) equation.

The free energy $F_{0}$ in (\ref{eq:feexpan}) is expressed as
\begin{equation}                                                \label{eq:FBP}
F_{0} = \sum\limits_{\gamma \in R} f_\gamma - \sum\limits_{(\mu,\nu)\in R}
f_{(\mu, \nu)} ,
\end{equation}
with
\begin{eqnarray}
f_\gamma  & \equiv    - \frac{1}{\beta} \ln\biggl[
\sum_{\underline{x}_\gamma}  \Psi_\gamma (\underline{x}_{\gamma})
      \prod_{\alpha \in \partial \gamma} p_{\alpha \rightarrow \gamma}
     (\underline{x}_{\gamma\cap\alpha}) \biggr] , \nonumber \\
f_{(\mu,\nu)}  & \equiv
 - \frac{1}{\beta} \ln \biggl[
 \sum_{\underline{x}_{\mu\cap\nu}}
     p_{\mu\rightarrow \nu}(\underline{x}_{\mu\cap\nu})
     p_{\nu \rightarrow \mu}(\underline{x}_{\mu\cap\nu})   \biggr] . \nonumber
\end{eqnarray}
The loop correction $L_{r}$ has the following expression
\begin{equation}\label{eq:correction}
    L_r =  \sum\limits_{\{\underline{x}_\gamma | \gamma \in r\}}
    \prod\limits_{\gamma \in r} w_\gamma(\underline{x}_\gamma)
     \prod_{(\mu, \nu) \in r} \Delta_{(\mu,\nu)} ,
\end{equation}
with
\begin{eqnarray}
    \Delta_{(\mu, \nu)} = \frac{e^{-y f_{(\mu,\nu)}}
     \prod_{i\in \mu\cap\nu}
     \delta( x_i^\mu, x_i^\nu)}
    {p_{\mu\rightarrow \nu}(\underline{x}_{\nu\cap\mu})
    p_{\nu\rightarrow \mu}(\underline{x}_{\mu\cap\nu})} - 1 , \nonumber \\
    w_\gamma(\underline{x}_\gamma) =
    \frac{\Psi_\gamma (\underline{x}_{\gamma})
      \prod_{\alpha \in \partial \gamma} p_{\alpha \rightarrow \gamma}
     (\underline{x}_{\gamma\cap\alpha})}
    {\sum_{\underline{x}_\gamma}  \Psi_\gamma (\underline{x}_\gamma)
      \prod_{\alpha \in \partial \gamma} p_{\alpha \rightarrow \gamma}
     (\underline{x}_{\gamma\cap\alpha})} . \nonumber
\end{eqnarray}

At a fixed point of the rgBP equation (\ref{eq:rgbp}), $F_{0}$
gives an approximate expression for the true free
energy of the system, when all the loop corrections in (\ref{eq:feexpan}) are
neglected. A nice property of $F_{0}$ is that, when viewed as a functional of
the probability functions
$\{p_{\mu\rightarrow \nu}, p_{\nu\rightarrow \mu}\}$, the
first variation of $F_{0}$ vanishes at a fixed point of (\ref{eq:rgbp}).
In other words, $F_{0}$ attains its extremal value at a rgBP fixed point.
Because of this extremal property, we regard a fixed point of (\ref{eq:rgbp})
as a macroscopic state of the model (\ref{eq:totalenergy}), with extremal free
energy value $F_{0}$. Using the counting number expression \eref{eq:rgcn}, we
can  re-express $F_0$ as
\begin{eqnarray}
F_0 &=& \sum\limits_{\gamma \in R} \Bigl[c_\gamma+ \sum_{\alpha > \gamma}
c_\alpha \Bigr] f_\gamma -\sum\limits_{(\mu\rightarrow \nu)}
\Bigl[c_\mu + \sum_{\alpha>\mu} c_\alpha \Bigr] f_{(\mu,\nu)} \nonumber \\
&= & \sum\limits_{\alpha\in R} c_\alpha \tilde{f}_\alpha ,
\label{eq:f02}
\end{eqnarray}
where $(\mu\rightarrow \nu)$ denotes a directed edge of $R$ from region $\mu$
to region $\nu$, and $\tilde{f}_\alpha$ is the free energy of region $\alpha$
as expressed by
\begin{equation}
\tilde{f}_\alpha =  \sum\limits_{\gamma \leq \alpha} f_\gamma
-\sum\limits_{(\mu,\nu) : \mu\leq \alpha, \nu \leq \alpha} f_{(\mu,\nu)} .
\label{eq:fa3}
\end{equation}
Notice (\ref{eq:f02}) has the same formal expression as the region-graph
variational free energy expression $F_{RG}$  of
\cite{Yedidia-Freeman-Weiss-2005}. But $F_0$ and $F_{RG}$ in general may not
be equivalent.

The statistical mechanics description of a single macroscopic state of model
(\ref{eq:totalenergy}) is composed of (\ref{eq:feexpan}) and (\ref{eq:rgbp}).
It can be regarded as forming a replica-symmetric spin-glass theory in the
region-graph representation. If the typical length of the loops of the
region-graph is long enough, since each $\Delta_{(\mu, \nu)}$ in the product
of \eref{eq:correction} has zero mean value under the probability measures
$\{w_\gamma(\underline{x}_{\gamma})\}$, the loop correction contributions to
the free energy might be negligible within each macroscopic state of the
system.

For the region-graph shown in \fref{fig:regiongraph} representing the the
square-lattice EA model, the explicit form of the rgBP equation \eref{eq:rgbp}
can be easily obtained. We have applied the rgBP equation on single instances
of the 2D EA model and obtained quite satisfactory results (will be reported
in a full paper). In the case that all the coupling constants
$J_{i j} \equiv J >0$ (the Ising model), the rgBP equation predicts a
phase-transition from the paramagnetic phase to the ferromagnetic phase at a
critical temperature value $T = 2.65635 J$. This value is lower than the value
of $T=2.8854 J$ as obtained by the Bethe-Peierls approximation, but still
higher than the exact critical temperature $T=2.2692 J$. Further improvements
on the critical temperature prediction should be achievable by working on
region-graphs with larger regions.


{\em Grand partition function expansion}---For graphical models with strong disorder
and frustration, very probably the rgBP equation (\ref{eq:rgbp}) has
multiple fixed points at low enough temperatures. This situation
corresponds to the existence of multiple low-temperature macroscopic states
$s$ of the model (\ref{eq:totalenergy}), each of which is characterized by
an extremal free energy value $F_{0}^{(s)}$. A full understanding of the
equilibrium property of the system then requires a quantitative description of
the free energy landscape at the level of macroscopic states. For this purpose,
we follow \cite{Xiao-Zhou-2011} (see also
\cite{Monasson-1995,Mezard-Parisi-2001}) and introduce a grand partition
function $\Xi(y; \beta)$
\begin{eqnarray}
\fl
\quad\quad
\Xi(y; \beta) &\equiv &  \sum_{s} e^{-y F_{0}^{(s)}} \nonumber \\
\fl
 &=& \prod\limits_{(\mu,\nu)\in R}
\int {\rm D} p_{\mu\rightarrow \nu} \int
{\rm D} p_{\nu\rightarrow \mu}
\delta\Bigl( p_{\mu\rightarrow \nu}-B_{\mu\rightarrow \nu}
\Bigr) \delta\Bigl( p_{\nu\rightarrow \mu}-B_{\nu\rightarrow \mu} \Bigr)
e^{-y F_{0}} .
\label{eq:Xi}
\end{eqnarray}
Because of the Dirac delta functions, only the rgBP fixed points
contribute to the grand partition function.
The parameter $y$ is the inverse temperature at the level of
macroscopic states.
For each edge $(\mu, \nu)$ of the
region-graph we introduce two auxiliary probability functions
$P_{\mu\rightarrow \nu}(p_{\mu\rightarrow \nu})$ and
$P_{\nu\rightarrow \mu}(p_{\nu\rightarrow \mu})$ and rewrite (\ref{eq:Xi}) as
\begin{eqnarray}
\fl \quad
\Xi(y; \beta) = \nonumber \\
\fl
\quad \quad \prod\limits_{\alpha \in R} \biggl[
\prod_{\gamma\in \partial \alpha} \int {\rm D} p_{\gamma\rightarrow \alpha}
P_{\gamma\rightarrow \alpha}(p_{\gamma\rightarrow \alpha}) e^{-y f_\alpha}
\biggr]  \prod\limits_{(\mu,\nu)\in R}
\biggl[
\frac{ \delta(p_{\mu\rightarrow \nu}-B_{\mu\rightarrow \nu}
) \delta(p_{\nu\rightarrow \mu}-B_{\nu\rightarrow \mu})}
{ e^{-y f_{(\mu,\nu)}} P_{\mu\rightarrow \nu}(p_{\mu\rightarrow \nu})
P_{\nu\rightarrow \mu}(p_{\nu\rightarrow \mu})}
\biggr] .
\label{eq:Xi-exp-2}
\end{eqnarray}

We can then follow \cite{Xiao-Zhou-2011} and obtain a loop series for the
expression (\ref{eq:Xi-exp-2}). The grand free energy of the system
$G(y; \beta) \equiv (-1/y) \ln \Xi(y; \beta)$ is written as
\begin{equation} \label{eq:gfeexpan}
G(y; \beta) = G_{0} - \frac{1}{\beta}
\ln \biggl[1+ \sum\limits_{r_{loop} \subseteq R}
L_{r_{loop}}^{(1)} \biggr] ,
\end{equation}
where the leading term $G_{0}$ has the expression
\begin{equation} \label{eq:gsp}
 G_{0}(y;\beta) = \sum\limits_{\gamma\in R} g_\gamma
 - \sum\limits_{(\mu,\nu)\in R} g_{(\mu,\nu)} ,
\end{equation}
with
\begin{eqnarray}
 g_\gamma &\equiv  - \frac{1}{y} \ln\biggl[
\prod_{\alpha \in \partial \gamma} \int {\rm D} p_{\alpha \rightarrow \gamma}
P_{\alpha \rightarrow \gamma}(p_{\alpha\rightarrow \gamma}) e^{-y f_\gamma}
\biggr] , \nonumber \\
g_{(\mu,\nu)} & \equiv  - \frac{1}{y}\ln\biggl[
\int \int {\rm D} p_{\mu\rightarrow \nu} {\rm D} p_{\nu\rightarrow \mu}
P_{\mu\rightarrow \nu}(p_{\mu\rightarrow \nu})
P_{\nu\rightarrow \mu}(p_{\nu\rightarrow \mu})
e^{-y f_{(\mu, \nu)}}
\biggr] . \nonumber
\end{eqnarray}
The correction contribution $L_r^{(1)}$ of a subgraph to the grand free energy
has a similar expression as (\ref{eq:correction}) \cite{Xiao-Zhou-2011}.
To ensure that any subgraph with at least one dangling edge has vanishing
correction contribution to the grand free energy, each probability function
$P_{\mu\rightarrow \nu} (p_{\mu\rightarrow \nu})$
needs to satisfy the following equation
\begin{equation} \label{eq:SPfix}
\fl
 P_{\mu\rightarrow \nu}(p_{\mu\rightarrow \nu})
 = S_{\mu\rightarrow \nu}[\{P_{\gamma\rightarrow \mu}\}]
\equiv
\frac{
\prod\limits_{\gamma\in \partial \mu\backslash \nu} \int
{\rm D} p_{\gamma\rightarrow \mu}
P_{\gamma\rightarrow \mu}(p_{\gamma \rightarrow \mu})
 e^{-y f_{\mu\rightarrow \nu}}
 \delta\bigl(p_{\mu\rightarrow \nu}-
B_{\mu\rightarrow \nu}\bigr)
}{
\prod_{\gamma\in \partial \mu\backslash \nu} \int
{\rm D} p_{\gamma\rightarrow \mu}
P_{\gamma\rightarrow \mu}(p_{\gamma\rightarrow \mu})
 e^{-y f_{\mu\rightarrow \nu}} } ,
\end{equation}
where
$$
f_{\mu\rightarrow \nu} \equiv
- \frac{1}{\beta} \ln\biggl[
\sum_{\underline{x}_\mu}  \Psi_\mu (\underline{x}_\mu)
      \prod_{\gamma \in \partial \mu \backslash \nu} p_{\gamma \rightarrow \mu}
     (\underline{x}_{\mu\cap\gamma}) \biggr] .
$$

Equation (\ref{eq:SPfix}) is called the region-graph survey-propagation (rgSP)
equation. If we neglect loop correction contributions, $G_{0}$ then gives an
approximate expression for the true grand free energy. By knowing the grand
free energy values at different inverse temperatures $y$, we can then calculate
the number $\exp[N \Sigma(f)]$ of macroscopic states with given extremal free
energy density $f\equiv F_{0}/N$ \cite{Mezard-Parisi-2001,Xiao-Zhou-2011}.
$\Sigma(f)$ is called the complexity function in spin-glass literature
\cite{Mezard-Montanari-2009}. It gives a quantitative description of the free
energy landscape of the $F_{0}$ functional (\ref{eq:FBP}). Notice that $G_{0}$
can again be viewed as a functional of the probability distributions
$\{P_{\mu\rightarrow \nu}, P_{\nu\rightarrow \mu}\}$. It is easy to check that
the first variation of $G_{0}$ vanishes at a fixed point of (\ref{eq:SPfix}),
suggesting that each rgSP fixed point corresponds to an extremal value of
$G_{0}$.

The statistical mechanics description of the model (\ref{eq:totalenergy}) at
the level of macroscopic states is composed of (\ref{eq:gfeexpan}) and
(\ref{eq:SPfix}). This description can be regarded as the first-step
replica-symmetry-breaking (1RSB) spin-glass theory in the region-graph
representation.

If the free energy functional $G_{0}$ has multiple extremal values, then a
higher-level partition function needs to be defined similar to (\ref{eq:Xi}),
and the free energy landscape of $G_{0}$ will then be described by a
higher-level message-passing equation. This hierarchy of partition function
expansions can be continued to even higher levels if necessary, leading to a
mathematical theory that can be regarded as the full-step RSB spin-glass
theory in the region-graph representation.

{\em Conclusion and discussions}---For a general graphical model (\ref{eq:totalenergy})
represented through a region-graph, a statistical mechanics theory was
rigorously derived in this work from the mathematical framework of
partition function expansion \cite{Xiao-Zhou-2011}. The obtained free energy
expressions and message-passing equations are exact in the sense the no
assumptions nor approximations were made during the theoretical construction.
This theory is most suitable for studying finite-dimensional models which
are full of short loops and strong local correlations. The theory reduces to
that of \cite{Xiao-Zhou-2011} in the special case of the region-graph being a
factor-graph.

Yedidia and co-authors \cite{Yedidia-Freeman-Weiss-2005} constructed a
variational free energy based on some heuristic arguments and consistency
requirements, and they then obtained several sets of generalized
belief-propagation equations. The starting point of the variational approaches
of \cite{Yedidia-Freeman-Weiss-2005} and \cite{Kikuchi-1951,An-1988} is the
non-equilibrium Gibbs free energy functional,  while our starting point is the
equilibrium partition function. Because of this qualitative difference, the
free energy expression $F_0$ in general may not be equivalent to the
variational free energy expression of \cite{Yedidia-Freeman-Weiss-2005}.
In an accompanying full paper we will investigate this issue more
systematically. A comparison between the rgBP (\ref{eq:rgbp}) and the
generalized belief-propagation schemes of \cite{Yedidia-Freeman-Weiss-2005}
will also be made. The possible links between our approach and the replica
cluster variation method of \cite{Rizzo-etal-2010} will also be thoroughly
discussed in the accompanying paper.

A distributed algorithm of message-passing can be implemented on
single finite-dimensional graphical models. To be efficient,
a suitable representation for the probability
distributions $p_{\mu\rightarrow \nu}(\underline{x}_{\nu\cap\mu})$
and $P_{\mu\rightarrow \nu}(p_{\mu\rightarrow \nu})$ should be worked out.
For the 2D square-lattice EA model \eref{eq:EAmodel}, if we represent it by
a region-graph shown in \fref{fig:regiongraph}, the rgBP equation can be
easily implemented. Systematic numerical results on the 2D and 3D EA model
with bimodal  coupling constants will be reported in another paper.

\ack

Support from the  "Collective Dynamics in Physical and Information Systems"
project of the Chinese Academy of Sciences is acknowledged.

\section*{References}


\begin{thebibliography}{10}

\bibitem{Mezard-etal-1987}
M.~M{\'{e}}zard, G.~Parisi, and M.~A. Virasoro.
\newblock {\em Spin Glass Theory and Beyond}.
\newblock World Scientific, Singapore, 1987.

\bibitem{Mezard-Montanari-2009}
M.~M{\'{e}}zard and A.~Montanari.
\newblock {\em Information, Physics, and Computation}.
\newblock Oxford Univ. Press, New York, 2009.

\bibitem{Mezard-etal-1986}
M.~M{\'{e}}zard, G.~Parisi, and M.~A. Virasoro.
\newblock {SK} model: the replica solution without replicas.
\newblock {\em Europhys. Lett.}, 1:77--82, 1986.

\bibitem{Monasson-1995}
R.~Monasson.
\newblock Structural glass transition and the entropy of the metastable states.
\newblock {\em Phys. Rev. Lett.}, 75:2847--2850, 1995.

\bibitem{Mezard-Parisi-2001}
M.~M{\'{e}}zard and G.~Parisi.
\newblock The bethe lattice spin glass revisited.
\newblock {\em Eur. Phys. J. B}, 20:217--233, 2001.

\bibitem{Mezard-etal-2002}
M.~M{\'{e}}zard, G.~Parisi, and R.~Zecchina.
\newblock Analytic and algorithmic solution of random satisfiability problems.
\newblock {\em Science}, 297:812--815, 2002.

\bibitem{Krzakala-etal-PNAS-2007}
F.~Krzakala, A.~Montanari, F.~{Ricci-Tersenghi}, G.~Semerjian, and
  L.~Zdeborova.
\newblock Gibbs states and the set of solutions of random constraint
  satisfaction problems.
\newblock {\em Proc. Natl. Acad. Sci. USA}, 104:10318--10323, 2007.

\bibitem{Edwards-Anderson-1975}
S.~F. Edwards and P.~W. Anderson.
\newblock Theory of spin glasses.
\newblock {\em J. Phys. F: Met. Phys.}, 5:965--974, 1975.

\bibitem{Belletti-Cotallo-Cruz-etal-2008}
F.~Belletti, M.~Cotallo, A.~Cruz, L.~A. Fernandez, A.~{Gordillo-Guerrero},
  M.~Guidetti, A.~Maiorano, F.~Mantovani, E.~Marinari, V.~{Martin-Mayor},
  A.~{Mu\~{n}oz Sudupe}, D.~Navarro, G.~Parisi, S.~{Perez-Gaviro}, J.~J.
  {Ruiz-Lorenzo}, S.~F. Schifano, D.~Sciretti, A.~Tarancon, R.~Tripiccione,
  J.~L. Velasco, and D.~Yllanes.
\newblock Nonequilibrium spin-glass dynamics from picoseconds to a tenth of a
  second.
\newblock {\em Phys. Rev. Lett.}, 101:157201, 2008.

\bibitem{Kikuchi-1951}
R.~Kikuchi.
\newblock A theory of cooperative phenomena.
\newblock {\em Phys. Rev.}, 81:988--1003, 1951.

\bibitem{An-1988}
G.~An.
\newblock A note on the cluster variation method.
\newblock {\em J. Stat. Phys.}, 52:727--734, 1988.

\bibitem{Yedidia-Freeman-Weiss-2005}
J.~S. Yedidia, W.~T. Freeman, and Y.~Weiss.
\newblock Constructing free-energy approximations and generalized
  belief-propagation algorithms.
\newblock {\em IEEE Trans. Inf. Theory}, 51:2282--2312, 2005.

\bibitem{Rizzo-etal-2010}
T.~Rizzo, A.~{Lage-Castellanos}, R.~Mulet, and F.~{Ricci-Tersenghi}.
\newblock Replica cluster variational method.
\newblock {\em J. Stat. Phys.}, 139:375--416, 2010.

\bibitem{Xiao-Zhou-2011}
J.-Q. Xiao and H.~Zhou.
\newblock Partition function loop series for a general graphical model:
  free-energy corrections and message-passing equations.
\newblock {\em J. Phys. A: Math. Theor.}, 44:425001, 2011.

\bibitem{Chertkov-Chernyak-2006b}
M.~Chertkov and V.~Y. Chernyak.
\newblock Loop series for discrete statistical models on graphs.
\newblock {\em J. Stat. Mech.: Theor. Exp.}, P06009, 2006.

\bibitem{Kschischang-etal-2001}
F.~R. Kschischang, B.~J. Frey, and H.-A. Loeliger.
\newblock Factor graphs and the sum-product algorithm.
\newblock {\em IEEE Trans. Inf. Theory}, 47:498--519, 2001.

\end{thebibliography}

\end{document}